\documentclass[prd,preprint,tightenlines,floatfix,showpacs,preprintnumbers,nofootinbib,eqsecnum,superscriptaddress]{revtex4}

\usepackage[dvips,final]{graphicx}
\usepackage{amssymb}
\usepackage{amsmath}
\usepackage{amsfonts}
\usepackage{epsfig}
\usepackage{bm}
\usepackage[section]{placeins}
\usepackage{multirow}
\usepackage{ctable}
\usepackage{booktabs}
\usepackage{array}
\usepackage{tabularx}
\usepackage{xcolor}
\usepackage{pstricks}
\usepackage{xspace}

\newcommand{\GeV}{\ensuremath{\mathrm{GeV}}\xspace}

\newcommand{\TeV}{\ensuremath{\mathrm{TeV}}\xspace}

\newcommand{\pperp}[1]{\ensuremath{\mathbf{p}_{#1\perp}}\xspace}
\newcommand{\kperp}[1]{\ensuremath{\mathbf{k}_{#1\perp}}\xspace}

\newcommand{\xbj}{\ensuremath{x_{\mathrm{Bj}}}\xspace}
\newcommand{\der}{\ensuremath{\mathrm{d}}\xspace}

\newcommand{\pythia}{PYTHIA 8\xspace}
\newcommand{\lpair}{LPAIR\xspace}

\begin{document}

\vfill
\title{Rapidity gap survival factors caused by remnant fragmentation for
 $W^+ W^-$ pair production via $\gamma^*\gamma^* \to W^+ W^-$
  subprocess with photon transverse momenta.}

\author{Laurent Forthomme}
\email{laurent.forthomme@cern.ch}
\affiliation{Institute of Physics and Astronomy, The University of Kansas, Lawrence, USA}
\altaffiliation{Now at Helsinki Institute of Physics, University of Helsinki, Finland}

\author{Marta {\L}uszczak}
\email{luszczak@ur.edu.pl}
\affiliation{
Faculty of Mathematics and Natural Sciences,
University of Rzesz\'ow, ul. Pigonia 1, PL-35-310 Rzesz\'ow, Poland}

\author{Wolfgang Sch\"afer}
\email{wolfgang.schafer@ifj.edu.pl}
\affiliation{Institute of Nuclear Physics Polish Academy of Sciences, ul. Radzikowskiego 152, PL-31-342 Krak\'{o}w, Poland}

\author{Antoni Szczurek}
\email{antoni.szczurek@ifj.edu.pl}
\affiliation{Institute of Nuclear Physics Polish Academy of Sciences, ul. Radzikowskiego 152, PL-31-342 Krak\'{o}w, Poland}

\date{\today}

\begin{abstract}
We calculate the cross section for
$p p \to W^{+} W^-$ in the recently developed
$k_{\rm T}$-factorisation approach, including transverse momenta of
the virtual photons.
We focus on processes with single and double proton dissociation.
First we discuss the gap survival on the parton level as due
to the emission of extra jet. Both the role of valence and sea
contributions is discussed.
The hadronisation of proton remnants is performed with {PYTHIA 8} string
fragmentation model, assuming a simple quark-diquark model for
proton. Highly excited remnant systems hadronise
producing particles that can be vetoed in the calorimeter.
We calculate associated effective gap survival factors.
The gap survival factors depend on the process, mass of the remnant
system and collision energy.
The rapidity gap survival factor due to remnant fragmentation
for double dissociative (DD) collisions ($S_{R,DD}$) is
smaller than that for single dissociative (SD) process ($S_{R,SD}$).
We observe the approximate factorisation $S_{R,DD} \approx (S_{R,SD})^2$,
however it is expected that this property will be violated by
soft rescattering effects not accounted for in this letter.

\end{abstract}

\pacs{}


\maketitle

\section{Introduction}

The processes with partonic $\gamma \gamma \to O_1 O_2$
($O_1$ and $O_2$ being electroweak states) subprocesses
become recently very topical.
Experimentally they can be separated from other competing processes
by imposing rapidity gaps around the electroweak vertex.
Both charged lepton pairs $l^+ l^-$
\cite{Chatrchyan:2011ci,Chatrchyan:2012tv,Aad:2015bwa,Cms:2018het,Aaboud:2017oiq}
and electroweak gauge bosons $W^+ W^-$ \cite{Khachatryan:2016mud,Aaboud:2016dkv}
were recently studied experimentally
at the Large Hadron Collider.
In particular processes with $W^+ W^-$ are of special interest as here
one can study e.g. anomalous quartic gauge boson couplings
\cite{Chapon:2009hh,Pierzchala:2008xc}.
Precise data may therefore provide a useful information allowing to test
the Standard Model in a sector, which is so far not accessible otherwise.

There are, in general, different categories of such processes depending
on whether the proton stays intact or undergoes an electromagnetic
dissociation (see e.g. \cite{daSilveira:2014jla,Luszczak:2015aoa}).

The $W^+ W^-$ production in proton-proton processes via the $\gamma \gamma \to W^+ W^-$
subprocess was recently studied in
collinear \cite{Luszczak:2014mta} and transverse momentum dependent factorisation
\cite{Luszczak:2018ntp} approaches.

Without additional requirements it is impossible to separate the
$\gamma \gamma \to W^+ W^-$ mechanism from $q \bar q \to W^+ W^-$,
$g g \to W^+ W^-$ or higher-order QCD processes.
To enhance the sample for the wanted mechanism one may impose a
rapidity gap condition around e.g. the  $e^+ \mu^-$ or $e^- \mu^+$ vertex in the leptonic
decay of the central diboson system.

\begin{figure}
  \centering
  \includegraphics[width=.32\textwidth]{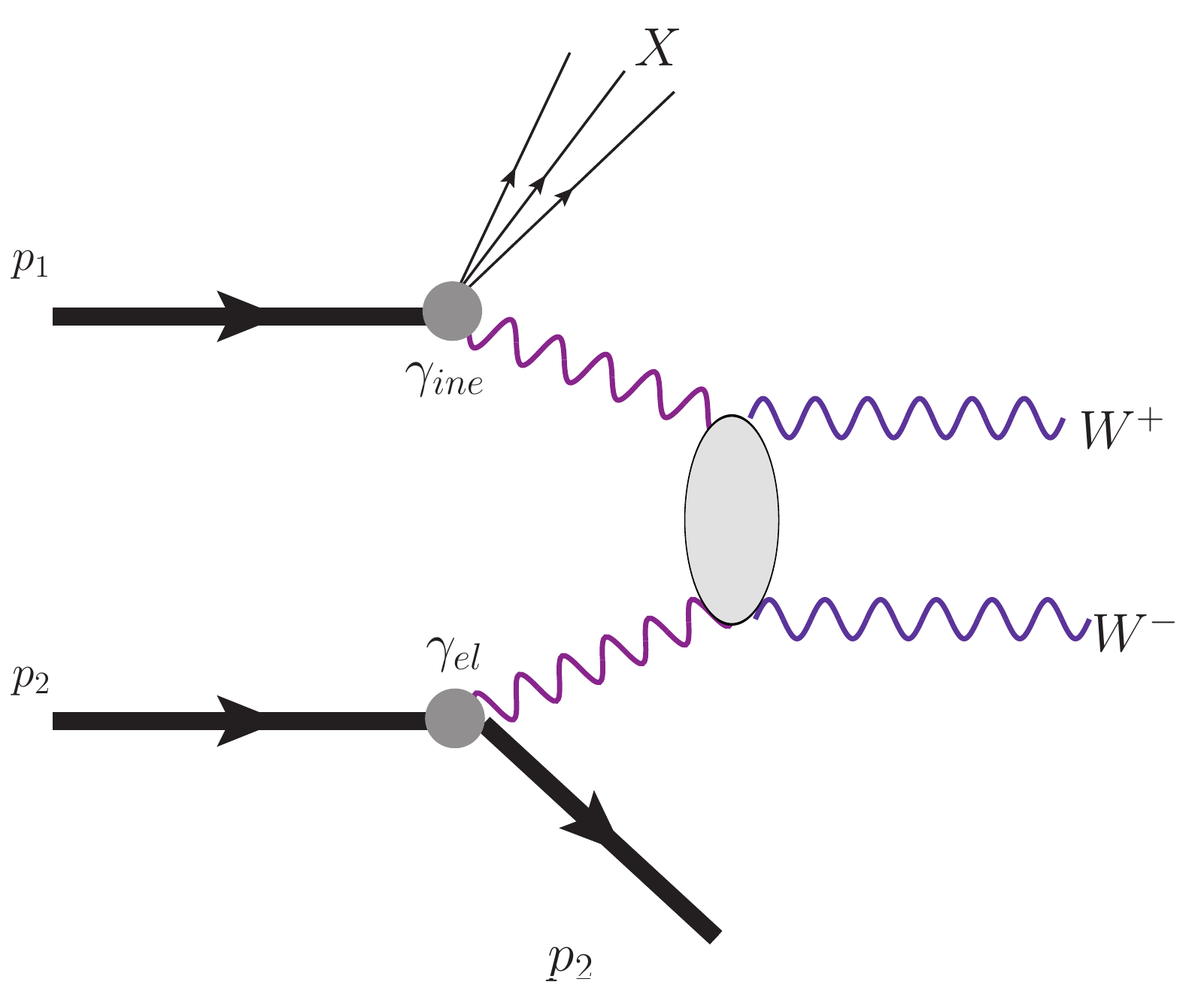}
  \includegraphics[width=.32\textwidth]{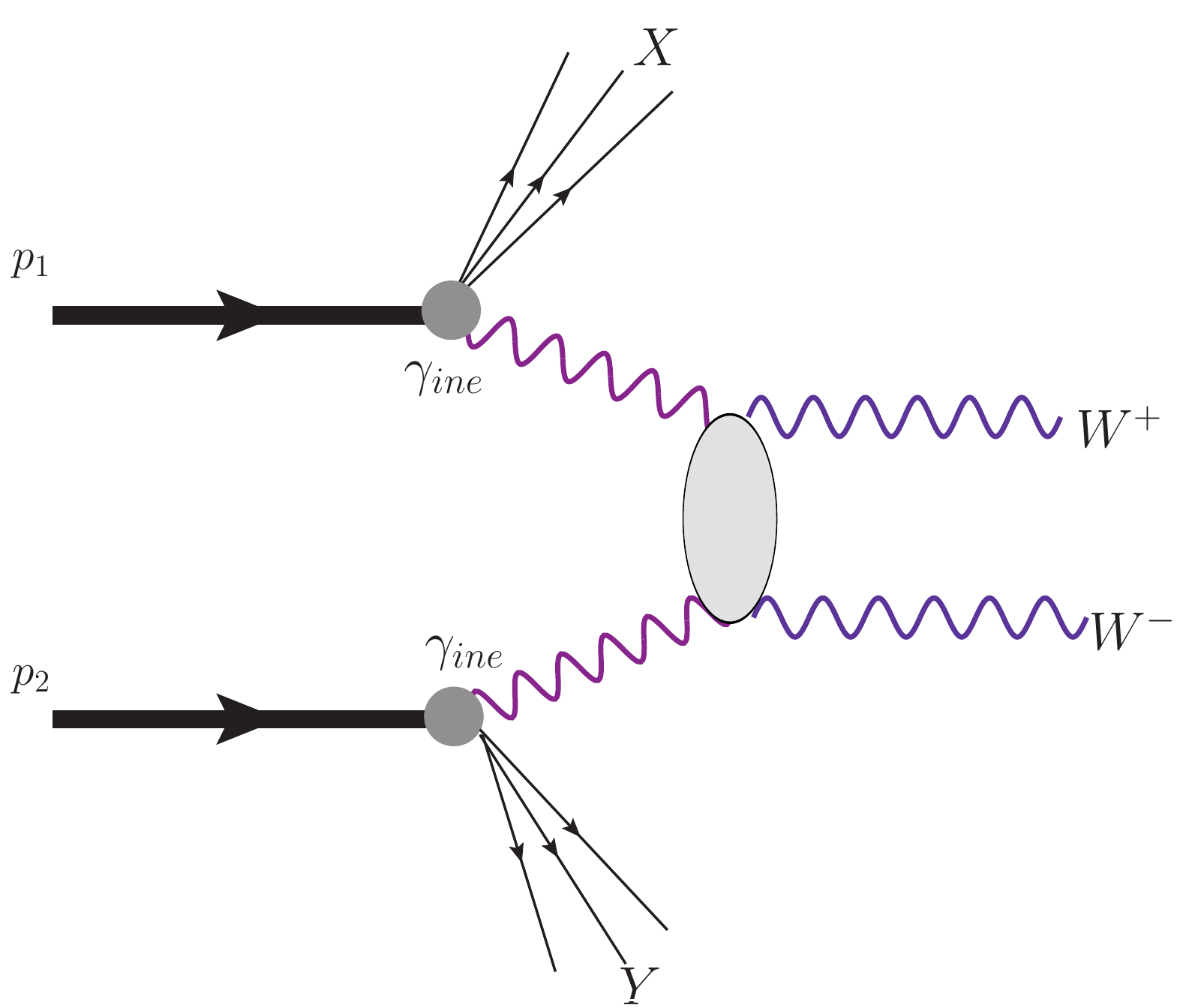}
  \caption{The single and double dissociative mechanisms discussed in the present letter.}
  \label{fig:diagram_DD}
\end{figure}

In Fig. \ref{fig:diagram_DD} we show a schematic picture of
the single and double dissociative two-photon processes.
In our recent paper \cite{Luszczak:2018ntp} we have shown that
rather large photon virtualities
and large mass proton excitation are characteristic for the
$\gamma \gamma \to W^+ W^-$ induced processes.
The highly excited hadronic systems hadronise producing (charged)
particles that may destroy the rapidity gap around the central event vertex.
The minimal requirement is to impose a condition of no charged particles
in the main ATLAS or CMS trackers.

We will focus on such effects in the present letter.
The hadronisation of the proton remnants will be performed
and conditions on charged particles will be imposed.
Our main aim is to estimate gap survival factor
associated with the remnant hadronisation, which destroys
the rapidity gap. Dependence on kinematic variables
will be studied.

As has been stressed in \cite{Harland-Lang:2016apc}, the ordinary
collinear photon parton distribution functions (PDFs) -- which imply a fully inclusive sum over
remnant final states -- cannot be used if additional gap requirements
are imposed on the final states.
For this purpose, in Ref. \cite{Harland-Lang:2016apc}
a concept of ``photon PDF in events with rapidity gaps'' was introduced.
There a requirement, that the parton emissions related to evolution
do not contaminate the central rapidity region is implemented.
The authors tried to approximately modify the collinear photon PDF
to include rapidity gap requirement(s) used in modern experiments.
The calculations in \cite{Harland-Lang:2016apc} are kept at the parton level, and no
explicit remnant hadronisation effects were discussed there.

The effect of gap survival related to the remnant
 fragmentation was discussed previously in the context of
the $e^+ e^-$ central production
in the framework of the \lpair code \cite{Baranov:1991yq}.

Remnant fragmentation is not the only effect that can destroy
the rapidity gap. There are also possible interactions between
the spectator partons of the colliding protons \cite{Bjorken:1992er}.
The gap survival factors for these processes are beyond the
scope of the present letter.
For recent estimates in photon induced processes, see e.g.
\cite{Dyndal:2014yea,Lebiedowicz:2015cea,Harland-Lang:2016apc}.
While \cite{Harland-Lang:2016apc} discusses results
at parton level it does include soft processes when calculating a
gap survival factor.

Within our approach the consistent inclusion of soft survival
effects remains a pressing issue for the future.

\section{Sketch of our calculational scheme}

We calculate cross section for the $p p \to W^+ W^-$ reaction
with double proton dissociation as:

\begin{eqnarray}
\frac{\der \sigma(pp \to X W^+ W^- Y)}{\der y_+ \der y_- \der^2\pperp{}^+ \der^2\pperp{}^- \der M_X \der M_Y} =
x_1 x_2 \int d^2 \kperp{1} d^2 \kperp{2}
\frac{\der \gamma(x_1,\kperp{1},M_X)}{\der M_X}
\frac{\der \gamma(x_2,\kperp{2},M_Y)}{\der M_Y} \; \times\hspace{5em}\nonumber \\
\times \; \frac{1}{16 \pi^2 (x_1 x_2 s)^2}
\sum_{\lambda_{W^+} \lambda_{W^-}} |M(\lambda_{W^+}, \lambda_{W^-}; \kperp{1},\kperp{2})|^2 \, \delta^{(2)}(\pperp{}^+ + \pperp{}^- - \kperp{1} - \kperp{2} ) 
.\nonumber \\ 
\end{eqnarray}
Here $y_\pm$ are the rapidities and $\pperp{}^\pm$ the transverse momenta of $W^\pm$ bosons.
The $M_X$-dependent photon fluxes can be decomposed into fluxes corresponding to
the relevant proton staying intact or dissociating (see Fig. \ref{fig:diagram_DD}):
\begin{eqnarray}
\frac{\der \gamma (x_1,\kperp{1},M_X)}{\der M_X} = \gamma_{\rm el}(x_1, \kperp{1}) \delta(M_X - m_p) +  \frac{\der \gamma_{\rm inel}(x_1,\kperp{1},M_X)}{\der M_X}
\theta\big(M_X -(m_p + m_\pi)\big),
\nonumber \\
\end{eqnarray}
and similarly for $(x_1, \kperp{1},M_X) \leftrightarrow (x_2, \kperp{2},M_Y)$,
so that the cross section for single dissociative process is less differential
as one of the two integrations over the remnant masses is unnecessary.
Such photon fluxes can be understood as a type of unintegrated parton distributions
\cite{Collins:2005uv}. They allow us to generate events containing remnants of mass $M_X, M_Y$.
Details on the relation of photon fluxes to proton structure functions and the used
matrix element $ M(\lambda_{W^+}, \lambda_{W^-}; \kperp{1}, \kperp{2})$ can be found in \cite{Luszczak:2018ntp} and
references therein.

Let us briefly recall the main ingredients for the construction of photon fluxes.
Elastic pieces only require the standard electromagnetic form factors of a proton.
The inelastic fluxes need the proton structure functions $F_2(\xbj,Q^2)$ and $F_L(\xbj, Q^2)$.
We use a parameterisation of $F_2$ and $F_L$,
which incorporates a large body of experimental data available
in different regions of $\xbj,Q^2$. For $Q^2>9 \, \rm GeV^2$ it uses a perturbative QCD NNLO calculation
\cite{Martin:2009iq}, while in the domain $Q^2< 9 \,\rm GeV^2$ the resonance region is described
by a fit found in \cite{Bosted:2007xd}, and elsewhere by the parameterisation of \cite{Airapetian:2011nu}.
For the longitudinal structure function, \cite{Abe:1998ym} is used to supplement \cite{Airapetian:2011nu}.

We use an implementation of the above process in CepGen \cite{Forthomme:2018ecc} for the
Monte-Carlo generation of unweighted events.

The hadronisation of remnant states $X$ and/or $Y$ is performed using the Lund
fragmentation algorithm implemented in \pythia \cite{Sjostrand:2014zea}, and interfaced to CepGen.
We model the incoming photon as emitted from a valence (up) quark collinear to the incoming proton direction.
Other flavour combinations are also expected to contribute to the process, but we observe
the kinematics of the outgoing $X$ and $Y$ systems is not sensitive to this choice.
The fractional quark momentum \xbj is determined event-by-event from the photon virtuality $Q^2$
and the relevant remnant mass $M_X$ through:
\begin{displaymath}
\xbj=\frac{Q^2}{Q^2+M_X^2-m_p^2}.
\end{displaymath}

We check the condition for each ``stable'' (pions, kaons, protons, \ldots)
charged particle produced in the hadronisation of $X$ and $Y$
remnants:
\begin{equation}
 -\eta_{\rm cut} < \eta^{\rm ch} < +\eta_{\rm cut} \; .
\label{charged_particle_condition}
\end{equation}
Each event for which at least one charged particle fulfils condition
(\ref{charged_particle_condition}) is discarded.
We introduce the ratio:
\begin{equation}
S_R({\omega}) = \frac{N_{\rm accepted}({\omega})}{N_{\rm all}({\omega})} \;,
\label{gap_survival_factor}
\end{equation}
where ${\omega}$ denotes a set of kinematic variables describing details
of the reaction.
$S_R({\omega})$ can be considered a phase-space-point-dependent rapidity gap
survival factor associated with remnant(s) fragmentation.
For example we will show such number for different ranges of masses of
the produced system both for double and single dissociation.

\section{Numerical results}

Here we wish to present some results of our Monte Carlo simulations.
We consider separately the case of double dissociation as well as
the case of single dissociation. The most important ingredient of our
calculation is a realistic hadronisation of proton remnants,
which allows to estimate the gap survival factor associated with
spoiling the rapidity gap in the central pseudorapidity region.
We assume a realistic situation $-2.5 < \eta < 2.5$, for individual (charged!)
particles, relevant for recent CMS \cite{Khachatryan:2016mud}
and ATLAS \cite{Aaboud:2016dkv} measurements.

It was shown e.g. in \cite{Luszczak:2018ntp} that without any gap
survival effects:
\begin{equation}
\sigma(\text{inel.-inel.}) > \sigma(\text{inel.-el.})+\sigma(\text{el.-inel.}) > \sigma(\text{el.-el.}) \; .
\label{inclusive_cs_ordering}
\end{equation}
Can this ordering be changed when the rapidity gap requirement is taken
into account?
As will be shown below, suppression effects
(due to emission of a jet and the remnant fragmentation)
are the biggest for inelastic-inelastic processes, so that in principle
the ordering in (\ref{inclusive_cs_ordering})
can be changed when a rapidity veto is imposed.

An important caveat has to be added: spectator parton rescatterings
can also change the hierachy of (\ref{inclusive_cs_ordering}).
Indeed, it is understood that these soft interactions will strongly
depend on the centrality of the collision in impact parameter space
\cite{Bjorken:1992er}. Photon exchange is generally long range
in impact parameter space, but events with large virtualities
$Q_{1,2}^2$ will be rather central and thus be more affected
by spectator rescatterings.


\subsection{Parton level approach for single dissociation}

Before studying the hadron level we wish to calculate the gap survival
factor on the parton level. In such a case it is the outgoing parton 
(jet or mini-jet),
which is struck by the virtual photon and destroys the rapidity gap.

\begin{figure}
\centering
\includegraphics[width=.63\textwidth]{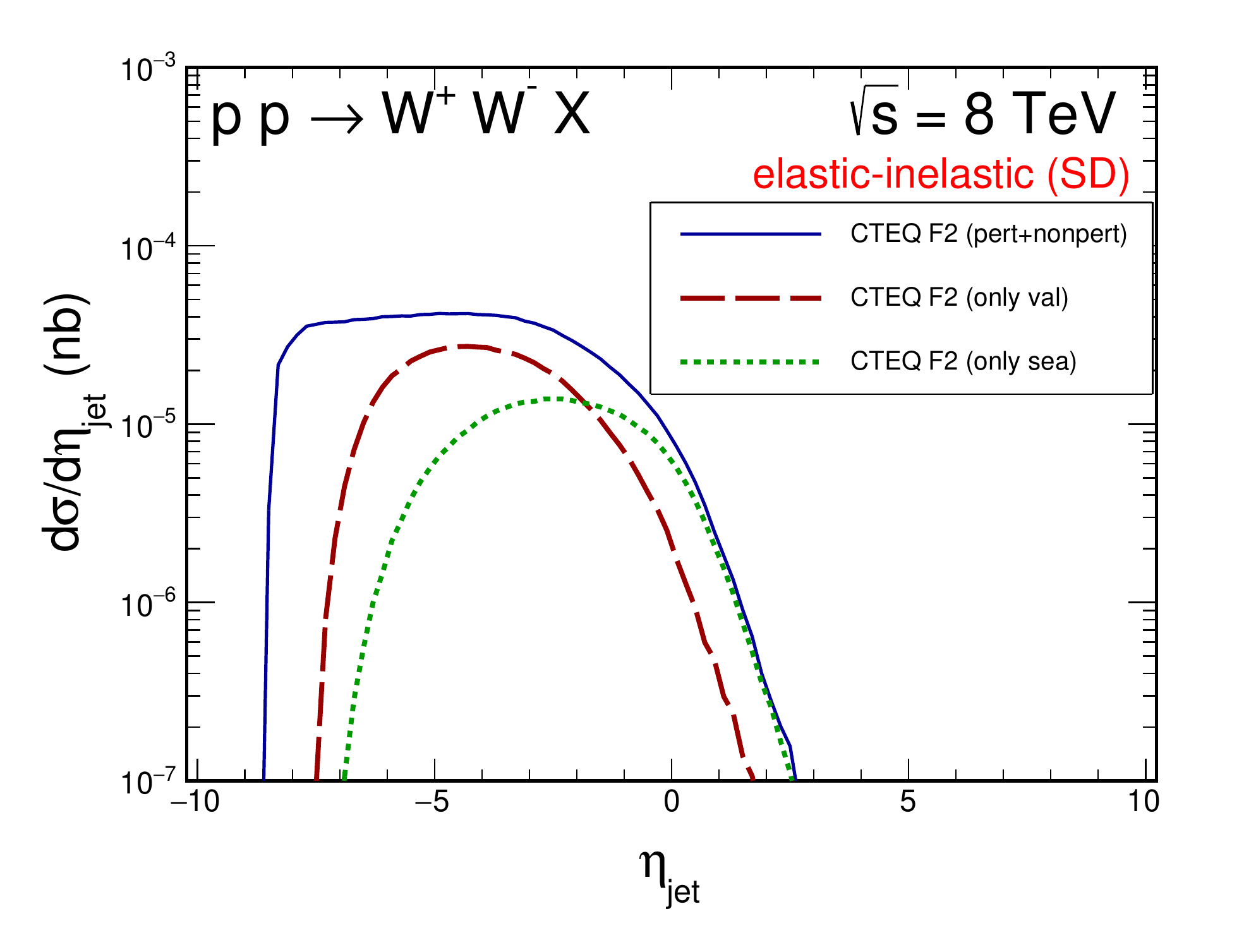}
\caption{Jet rapidity distribution for $F_2$ using a LO
partonic distribution at large $Q^2$. The solid line is a sum of all contributions.
The dashed line is for the valence component and the dotted line is for
the sea component.
}
 \label{fig:dsigma_dyjet}
\end{figure}

The gap survival factor can be then defined as:
\begin{equation}
S_R(\eta_{\rm cut}) = 1 - {1 \over \sigma}
\int_{-\eta_{\rm cut}}^{\eta_{\rm cut}} \frac{{\rm d}\sigma}{{\rm d}
    \eta_{\rm jet}} {\rm d} \eta_{\rm jet}, \;
\label{parton_model_gap_survival}
\end{equation}
where ${\rm d}\sigma / {\rm d}\eta_{\rm jet}$ is the rapidity distribution of the cross
section for $W^+ W^-$ production as a function of rapidity of the extra
jet ({\emph de facto} parton) and $\sigma$ is the associated integrated cross
section.
In Fig. \ref{fig:dsigma_dyjet} we show ${\rm d}\sigma /{\rm d}\eta_{\rm jet}$ as a function
of $\eta_{\rm jet}$. No extra cuts are imposed here.
We get a very broad distribution in $\eta_{\rm jet}$ (see solid line).

Different processes contribute to this distribution: the jet may originate
from the valence or sea (anti-)quark distribution.
Unfortunately the NNLO calculation \cite{Martin:2009iq} does not allow a
straightforward decomposition into sea and valence.

However, a leading-order parameterisation may be used to disentangle the partonic contribution to
$F_2(\xbj,Q^2)$ for $Q^2 > 9 \, \rm GeV^2$.
For illustration we show the contributions of valence (dashed line)
and sea (dotted line) components in Fig. \ref{fig:dsigma_dyjet}.
As can be seen, rapidity distributions for different components are very different. The
sea component is important for larger rapidities than the valence one.
There is also a non-perturbative component at very negative rapidities.
Notice that we generate events which include remnants of masses 
$M_X,M_Y$. All information on the excitation of these states is encoded
in the proton structure functions $F_2, F_L$, taken essentially from data. 
In particular, it includes the excitation
of baryon resonances and low-mass hadronic continua.
The regions of low values of $(Q_1^2)$ and/or $Q_2^2)$ are called
nonperturbative as there the partonic picture is not enough.
Our approach to the final state is very different from the one using 
collinear factorized partons, where an inclusive sum over all baryon
remnants is implied. It is only after such an inclusive sum 
that the nonperturbative contribution at low $Q^2$ in our
parametrization could be reexpressed in
terms of the initial condition of ``DGLAP'' photons. 
Clearly this is not useful for the problem at hand.

\begin{figure}
\centering
\includegraphics[width=.63\textwidth]{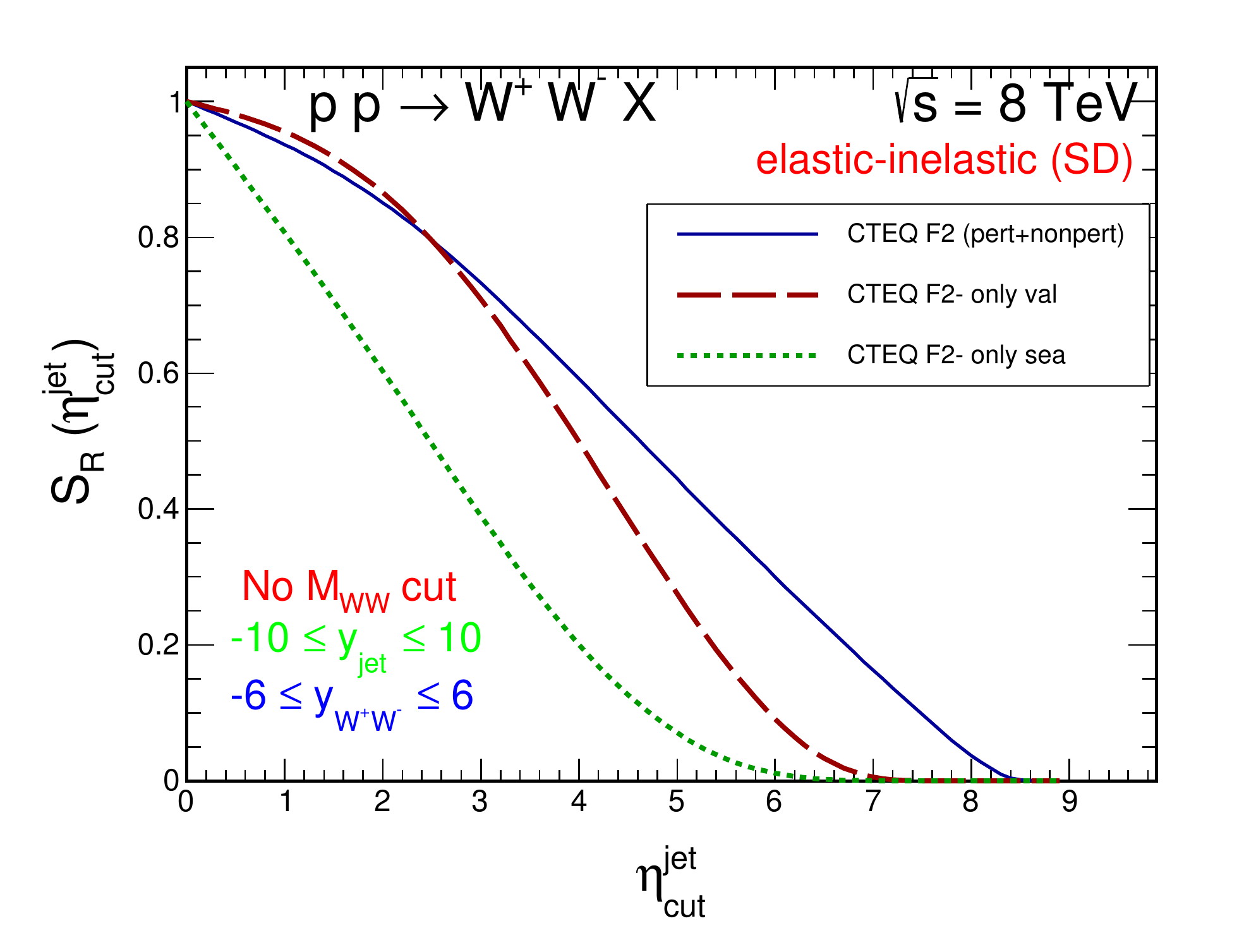}
\caption{Gap survival factor associated with the jet emission and
defined by Eq. \eqref{parton_model_gap_survival}. The solid line is for
the full model, the dashed line for the valence contribution
and the dotted line for the sea contribution.
}
 \label{fig:S_R_partonic}
\end{figure}

Now we shall present the parton level gap survival factor as a function of
the somewhat artificial window $(-\eta_{cut}, \eta_{cut})$ which is free
of the outgoing parton (jet).
We show corresponding $S_R(\eta_{cut})$ in Fig. \ref{fig:S_R_partonic}.
The solid line represents our partonic result. For comparison we show also
$S_R$ when only one component (valence or sea) of $F_2$ is included
in the calculation, see dashed and dotted lines. 
In this case, the cross section $\sigma$ in
the denominator of Eq.~\eqref{parton_model_gap_survival} is the integral
of the relevant component (sea or valence) only. 
We see that gap survival factors for the different components are 
fairly different.
Our final result (solid line) correctly includes all components. 
Please notice that according to Eq.~\eqref{parton_model_gap_survival} 
the solid line is not the sum of the dashed and dotted curves. 
The distribution of $S_R$ for the full model (solid curve) extends 
to much larger $\eta_{\rm cut}$ than the valence and sea 
contributions separately. This is due to a nonperturbative contribution
(see a comment above), which dominates at very large
negative rapidities (see the $\eta_{\rm jet}$ distribution in 
Fig.~\ref{fig:dsigma_dyjet} ). The emitted jets can be associated
only with partonic component of the model structure function. 

\subsection{Particles in the jet}
\begin{figure}
\includegraphics[width=.6\textwidth]{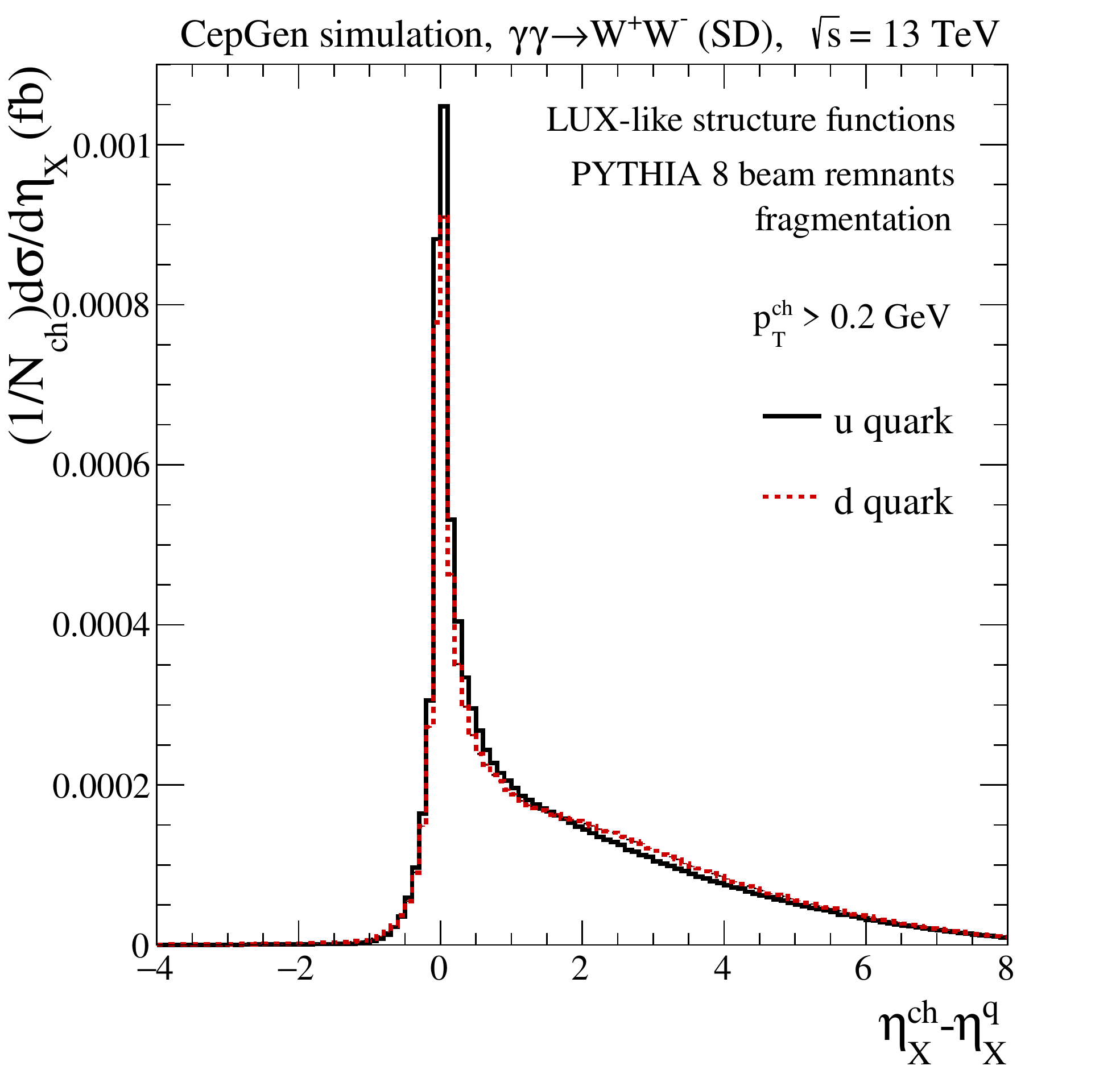}
\caption{
Distribution of charged particles in the single dissociative case for $u$ (black solid line)
and $d$ (red dotted line) quarks at $13~\TeV$ with respect to $\eta_{\rm jet}$.
}
\label{fig:y_in_thejet}
\end{figure}

Now we wish to show pseudorapidity distribution of charged particles
relative to the parton (jet) rapidity
($\Delta \eta = \eta_{\rm ch} - \eta_{\rm jet}$). 
In Fig. \ref{fig:y_in_thejet} we see a sharp peak relative to zero
which can be interpreted as the distribution within the jet. To the right of the peak
we see in addition the contribution of beam remnants which leads to a visible asymmetry
of the distribution.
This means that the corresponding gap survival factor should be very similar
when using the particle closest in rapidity space to the central system as that when using
(pseudo)rapidity of the jet (parton).
The effect of hadronisation is an order of magnitude smaller than
the effect for different components (valence, sea, etc.).

\subsection{Double dissociation}

\begin{figure}
\includegraphics[width=.48\textwidth]{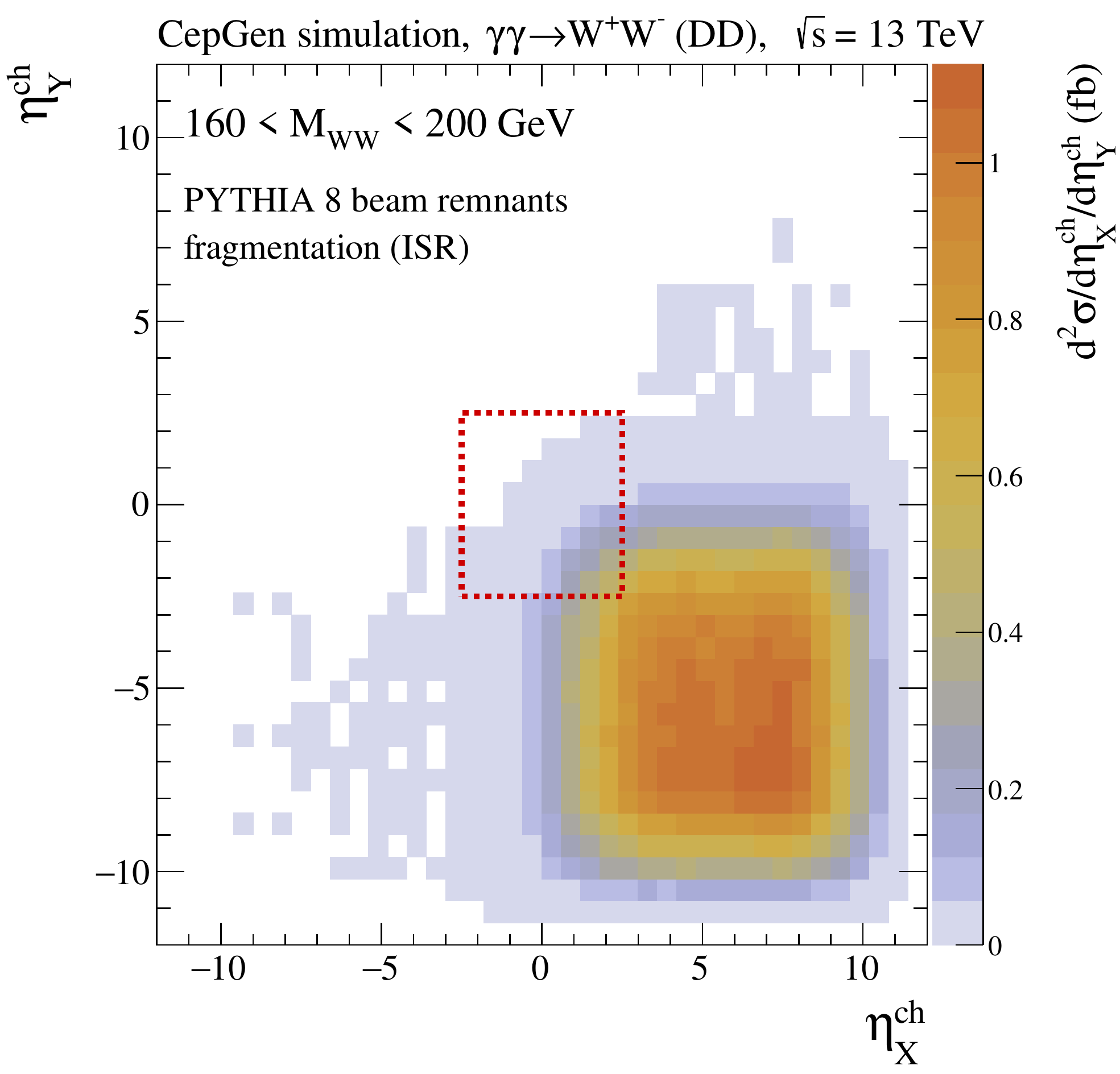}
\includegraphics[width=.48\textwidth]{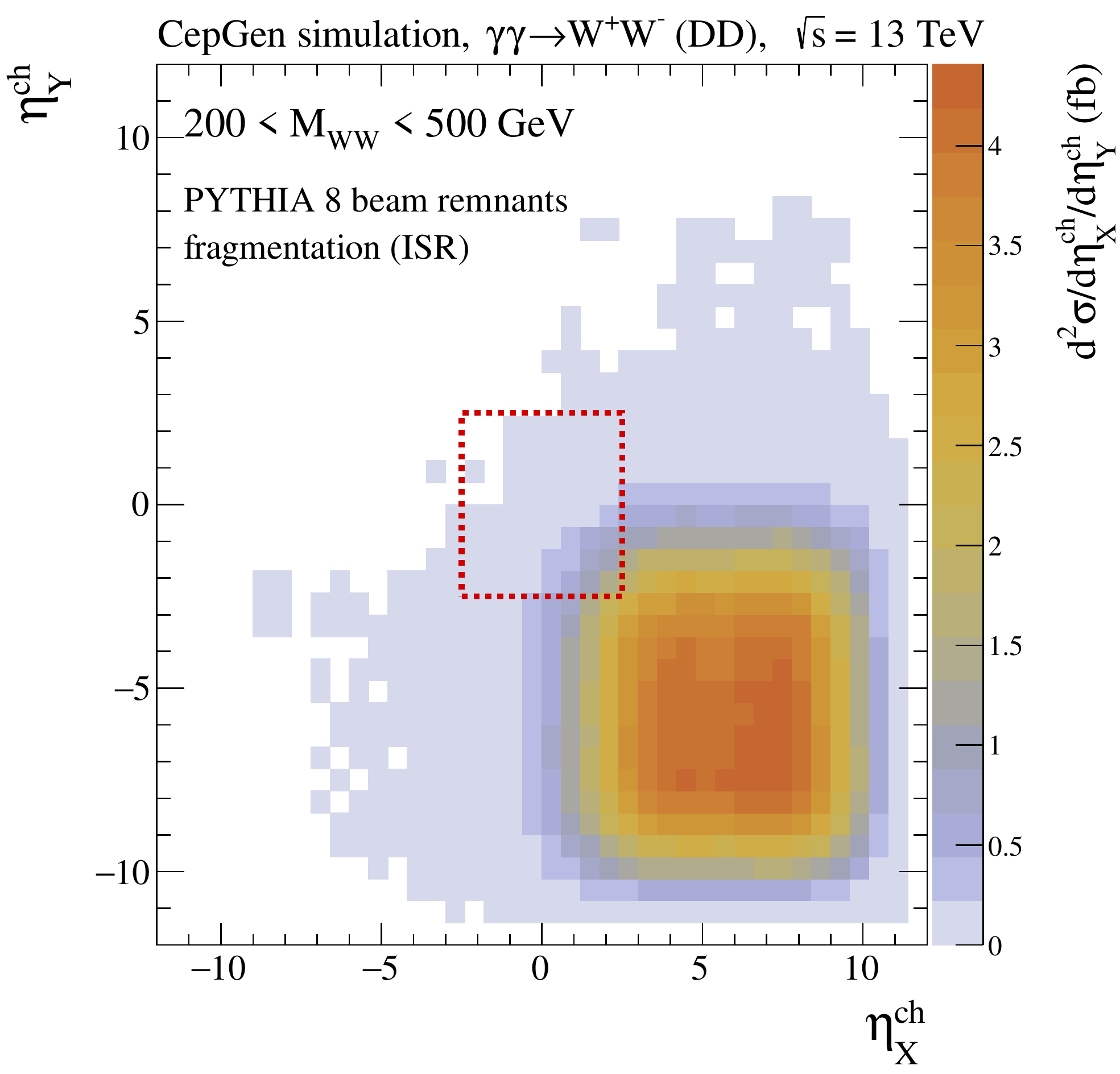}\\
\includegraphics[width=.48\textwidth]{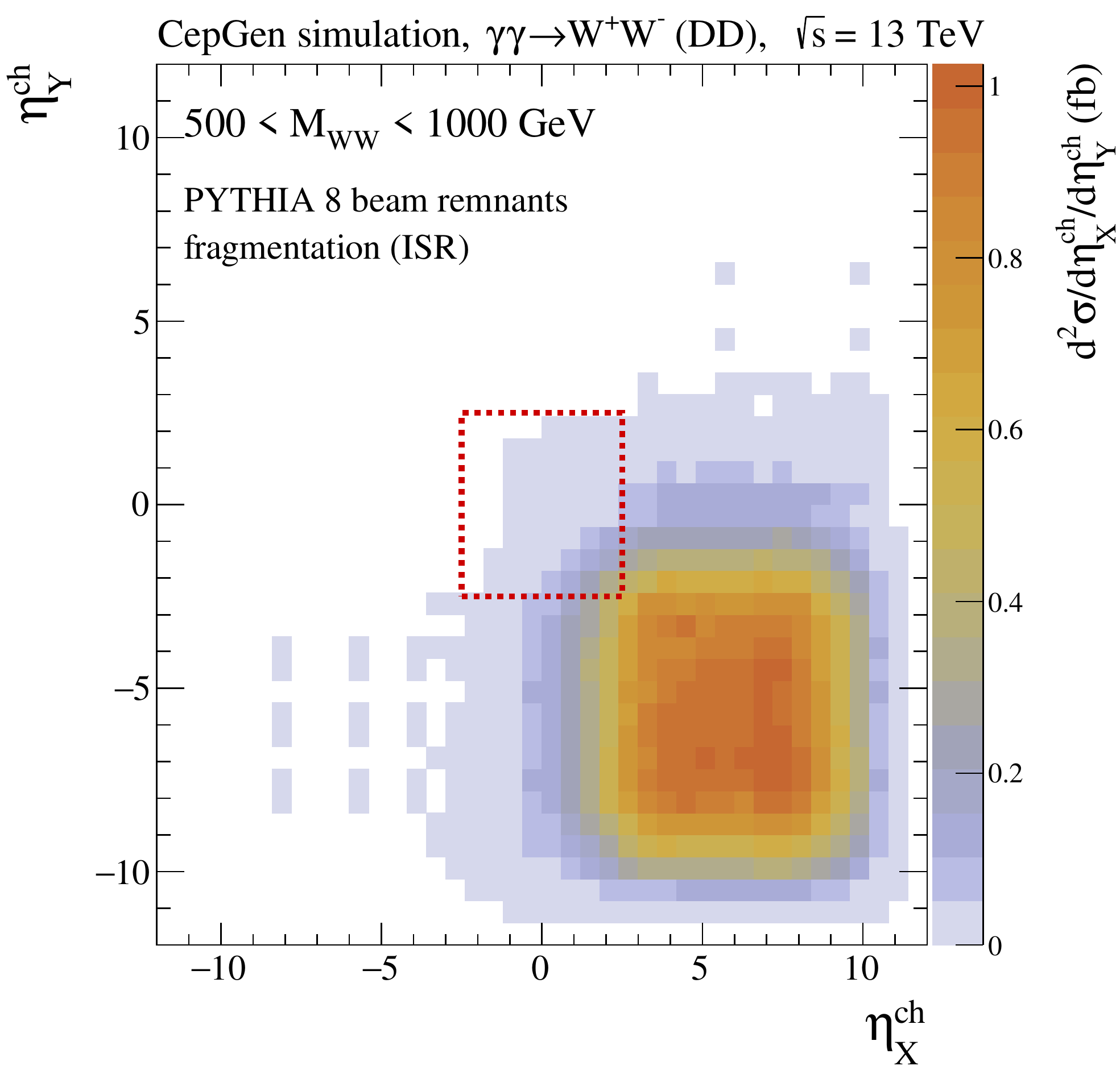}
\includegraphics[width=.48\textwidth]{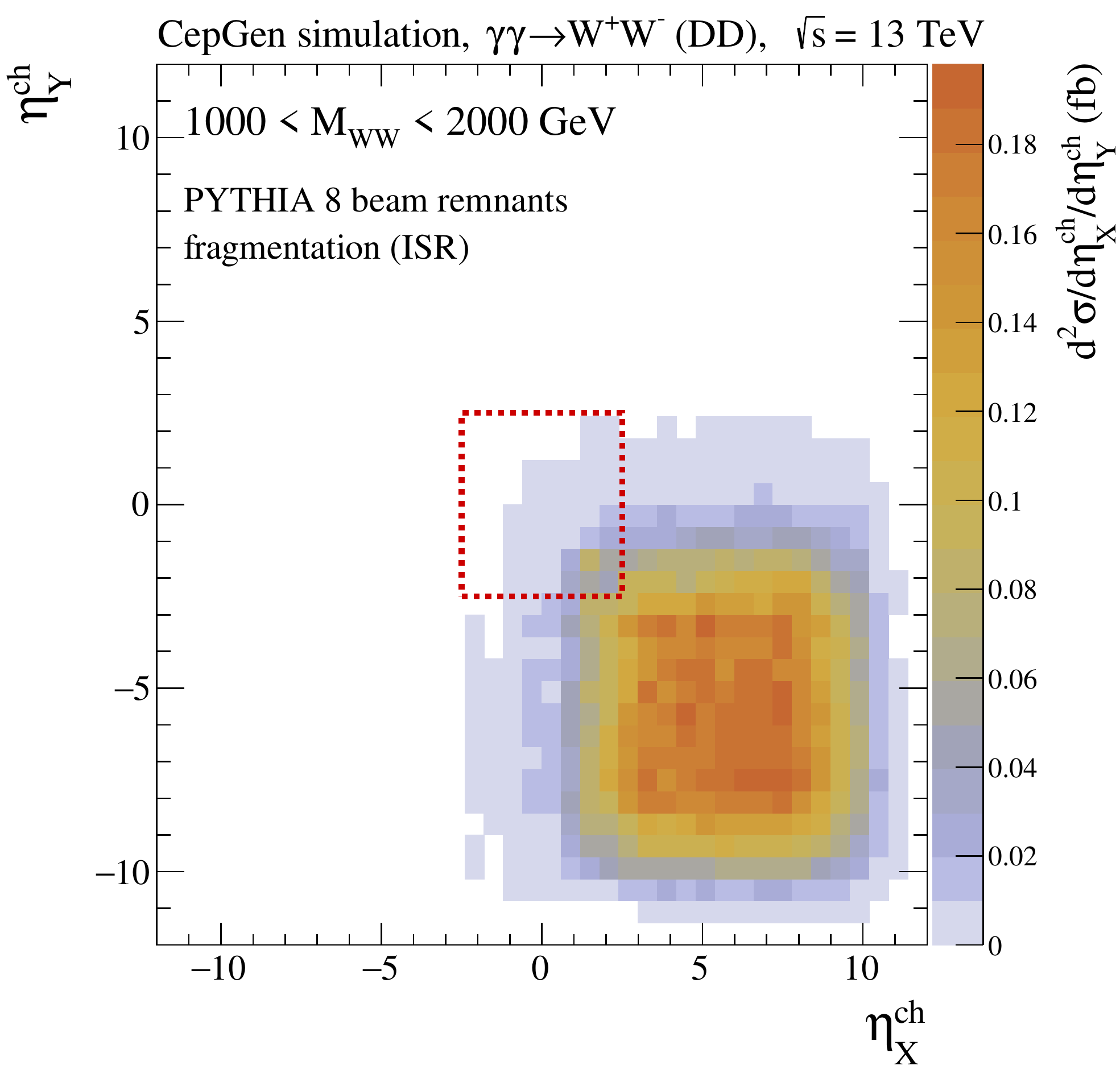}
\caption{Two-dimensional ($\eta^{\rm ch}_{X},\eta^{\rm ch}_{Y}$) distribution
for four different windows of $M_{WW}$: $(2 M_W, 200~\GeV)$,
$(200, 500~\GeV)$, $(500, 1000~\GeV)$, $(1000, 2000~\GeV)$.
The square shows pseudorapidity coverage of ATLAS or CMS inner tracker.}
\label{fig:dsig_deta1deta2_MWW_windows}
\end{figure}

We start the detailed studies on the hadron level
(including hadronisation)
from the largest contribution, in the inclusive case, the inelastic-inelastic
(double dissociative) \cite{Luszczak:2018ntp} processes.
In this case both remnants fragment and we have to include their
fragmentation simultaneously.
In Fig. \ref{fig:dsig_deta1deta2_MWW_windows} we show two-dimensional
distributions in pseudorapidity of particles from $X$ ($\eta^{\rm ch}_X$)
and $Y$ ($\eta^{\rm ch}_Y$) for different ranges of masses of the centrally
produced system. For illustration the region relevant for ATLAS and CMS
pseudorapidity coverage is pictured by the thin dashed square.

The two dimensional plots are not sufficient to see
a dependence of the associated gap survival factor on the mass of
the centrally produced system.

We quantify this effect, see Table \ref{tab:gsf_sd}, by showing average remnant rapidity
gap factors for different ranges of $M_{WW}$ masses.
There we observe a mild dependence.
The remnant rapidity gap survival factor at fixed $\eta_{\rm cut}$
becomes larger at higher collision energies.

\begin{table}[tbp]
\centering
\begin{tabular}{|c|c|c|c|c|c|c|}
\hline
\multirow{2}{*}{Contribution} & \multicolumn{2}{c|}{$S_{R,SD}(|\eta^{\rm ch}|<2.5)$ } & \multicolumn{2}{c|}{$\left(S_{R,SD}\right)^2(|\eta^{\rm ch}|<2.5)$ } & \multicolumn{2}{c|}{$S_{R,DD}(|\eta^{\rm ch}|<2.5)$}\\
\cline{2-7}
                       & $8~\TeV$  & $13~\TeV$ & $8~\TeV$  & $13~\TeV$  & $8~\TeV$ & $13~\TeV$\\
\hline
$(2 M_{WW}, 200~\GeV)$ & 0.763(2)  & 0.769(2)  & 0.582(4)  & 0.591(4)   & 0.586(1) & 0.601(2)\\
\hline
$(200, 500~\GeV)$      & 0.787(1)  & 0.799(1)  & 0.619(2)  & 0.638(2)   & 0.629(1) & 0.649(1)\\
\hline
$(500, 1000~\GeV)$     & 0.812(2)  & 0.831(2)  & 0.659(3)  & 0.691(3)   & 0.673(2) & 0.705(2)\\
\hline
$(1000, 2000~\GeV)$    & 0.838(7)  & 0.873(5)  & 0.702(12) & 0.762(8)   & 0.697(5) & 0.763(6)\\
\hline
full range             & 0.782(1)  & 0.799(1)  & 0.611(2)  & 0.638(2)   & 0.617(1) & 0.646(1)\\
\hline
\end{tabular}
\caption{Average rapidity gap survival factor related to remnant fragmentation
	for {\it single dissociative} and {\it double dissociative} contributions
for different ranges of $M_{WW}$.
All uncertainties are statistical only.
}
\label{tab:gsf_sd}
\end{table}


\begin{figure}
  \centering
  \includegraphics[width=.48\textwidth]{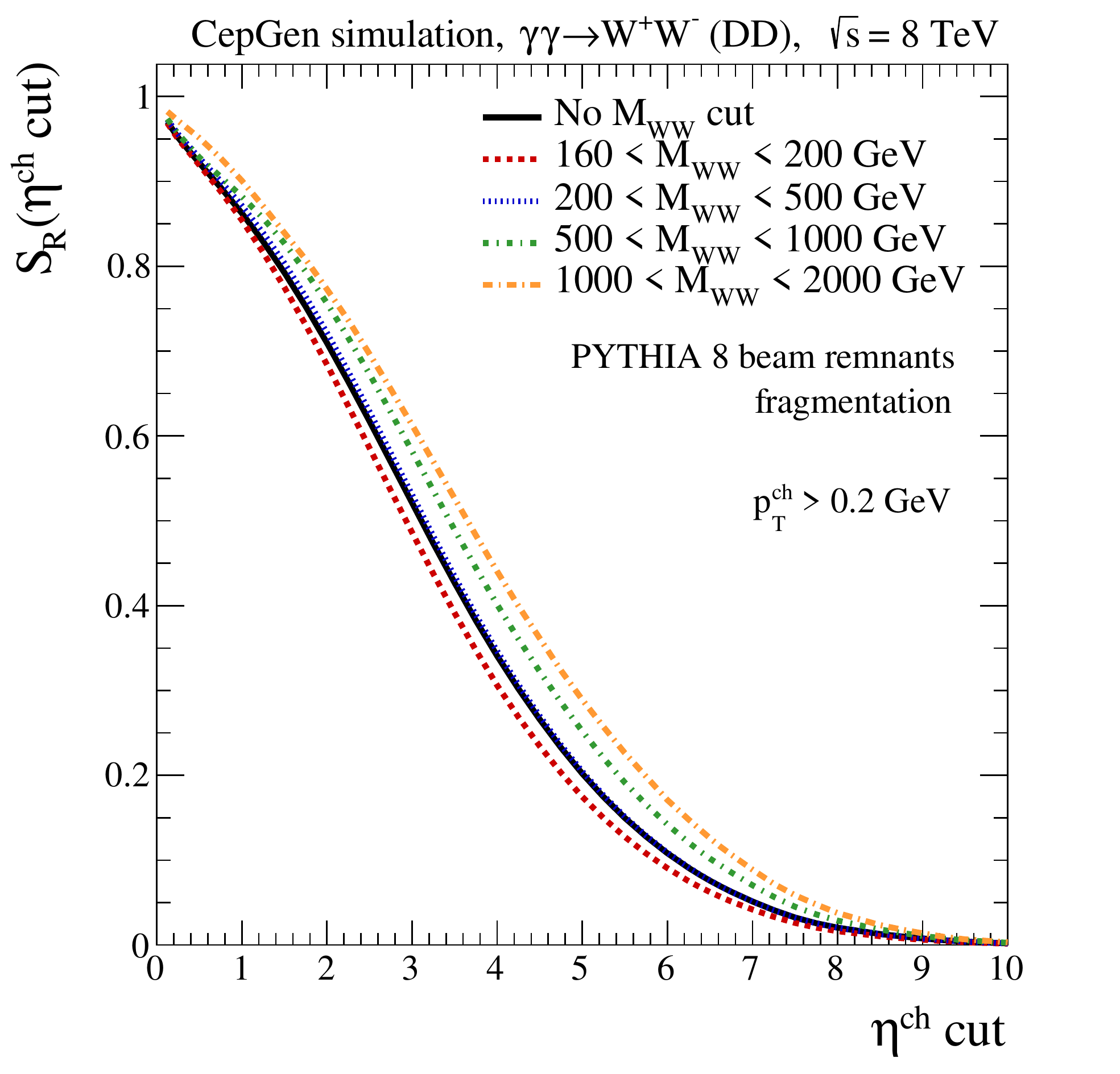}
  \includegraphics[width=.48\textwidth]{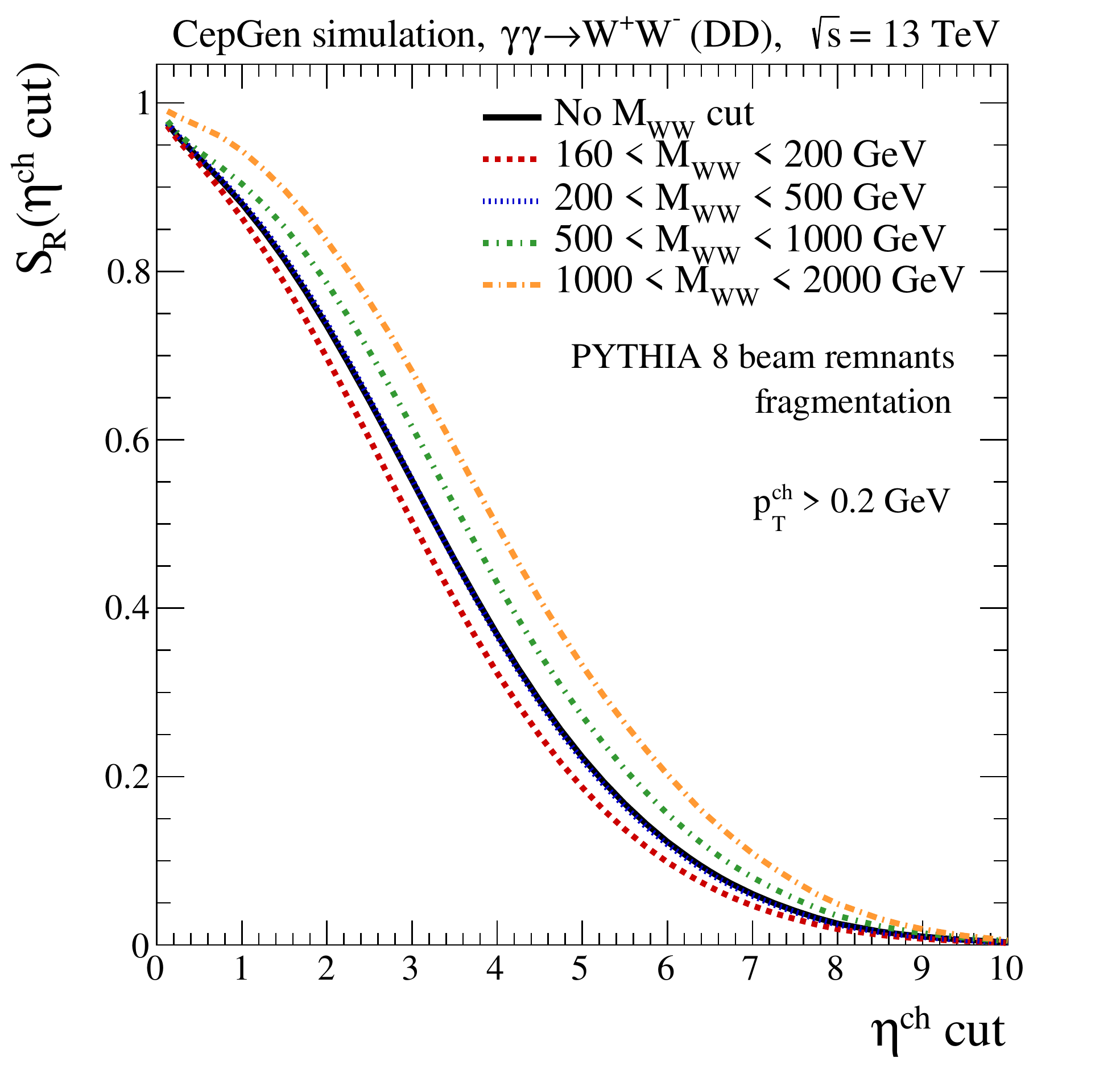}
  \caption{Gap survival factor for double dissociation as a function of
    the size of the pseudorapidity veto applied on charged particles emitted from
    proton remnants, for the
    diboson mass bins defined in the text and in the figures for $\sqrt{s} = 8~\TeV$ (left)
    and $13~\TeV$ (right).}
  \label{fig:surv_vs_etacut_dd}
\end{figure}
In Fig. \ref{fig:surv_vs_etacut_dd} we show the distribution in $\eta_{\rm cut}$ for the double
dissociation process. We predict a strong dependence on $\eta_{\rm cut}$. It would be
valuable to perform experimental measurements with different $\eta_{\rm cut}$.

\subsection{Single dissociation}

\begin{figure}
  \centering
  \includegraphics[width=.48\textwidth]{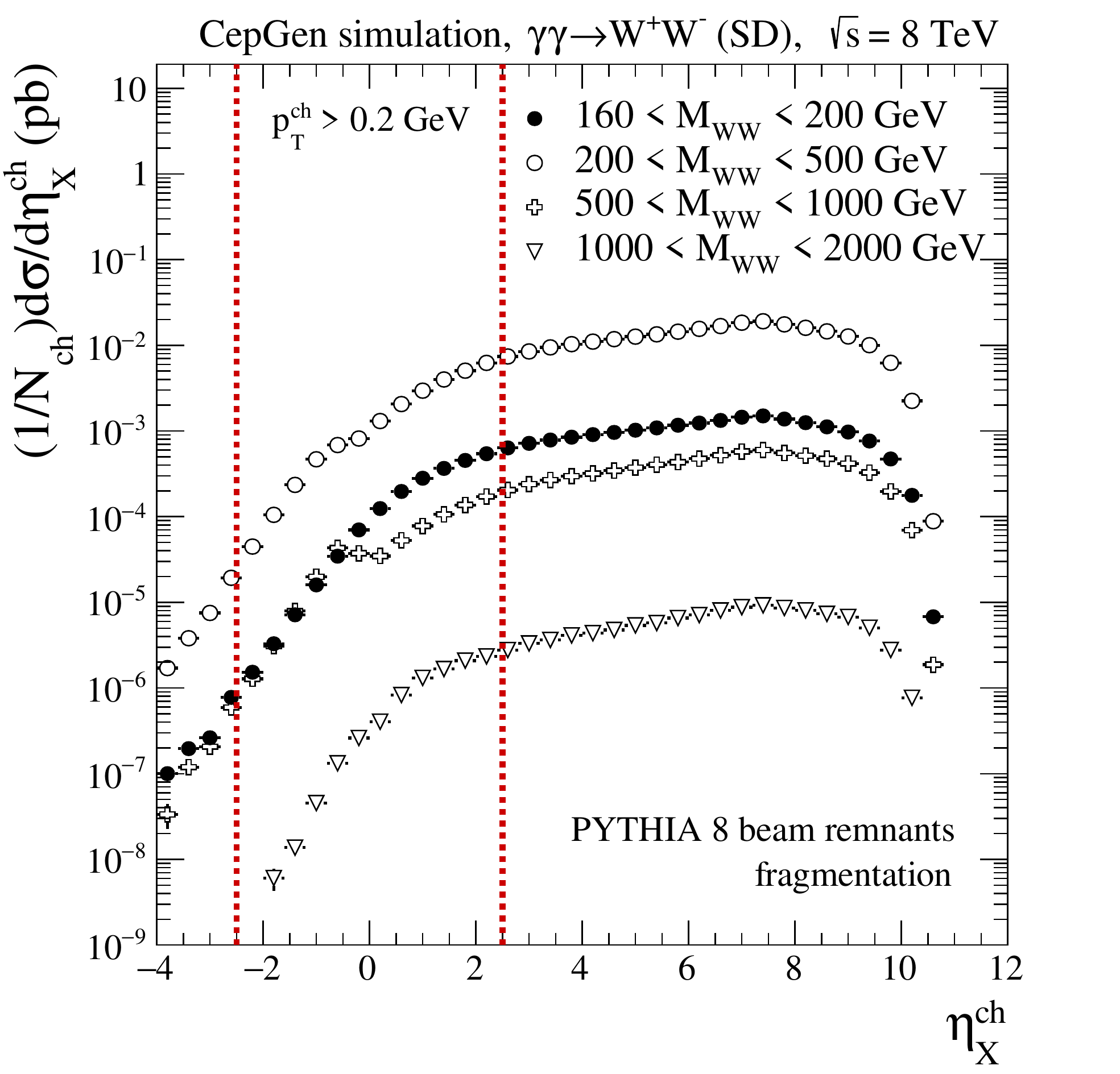}
  \includegraphics[width=.48\textwidth]{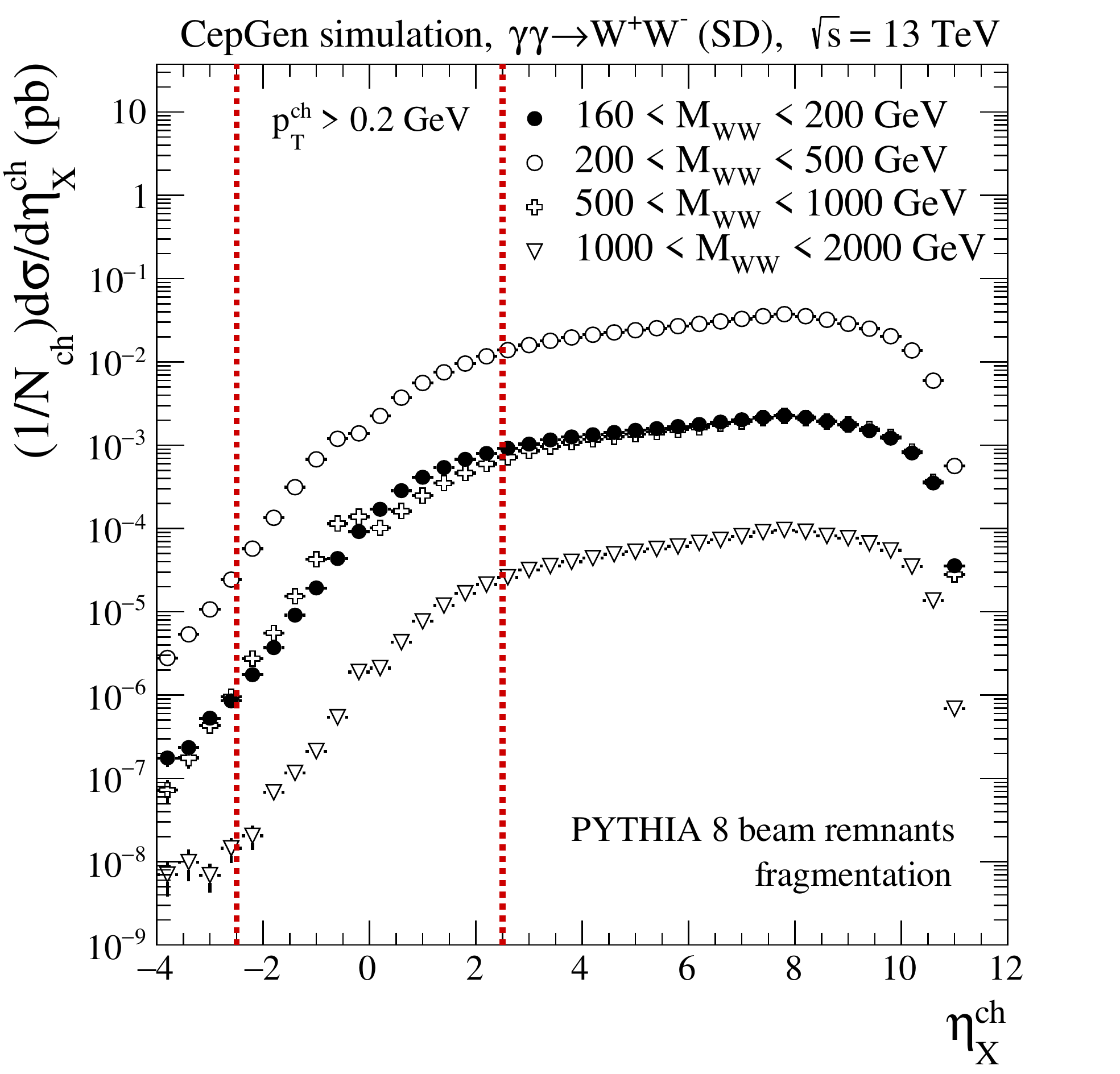}
  \caption{$\eta_{\rm ch}$ distribution for single dissociative process
  for four different windows of $M_{WW}$: $(2 M_W, 200~\GeV)$,
  $(200, 500~\GeV)$, $(500, 1000~\GeV)$, $(1000, 2000~\GeV)$, and for
  $\sqrt{s} = 8~\TeV$ (left) and $13~\TeV$ (right).
  The lines show pseudorapidity coverage of ATLAS or CMS detector.}
  \label{fig:dsig_deta_MWW_windows}
\end{figure}

We repeat a similar analysis for the single dissociative process.
In Fig. \ref{fig:dsig_deta_MWW_windows} we show the rapidity distribution
of charged particles produced in the fragmentation of the $X$ system.
The contamination of the detector is only weakly
correlated with the mass of the centrally produced system.

Again we quantify the effect by showing the average remnant rapidity gap
survival factor for the same windows of $M_{WW}$.
The conclusions here are similar as for the double dissociation,
except that the effect of destroying the rapidity gap is smaller.

In Table \ref{tab:gsf_sd} we show the rapidity gap survival factor for single
and double dissociation processess. The middle column shows the square of single
dissociation survival factors.
By comparing the latter results with the ones
for double dissociation, collected in Table \ref{tab:gsf_sd} we
observe that with good precision:

\begin{equation}
S_{R,DD} \approx \left(S_{R,SD}\right)^2  \; .
\label{factorisation}
\end{equation}
Such an effect is naively expected when the two fragmentations
are independent, which is the case by the model construction.
Again, we repeat the caveat, that soft processes will violate
the factorisation discussed here.

\begin{figure}
  \centering
  \includegraphics[width=.48\textwidth]{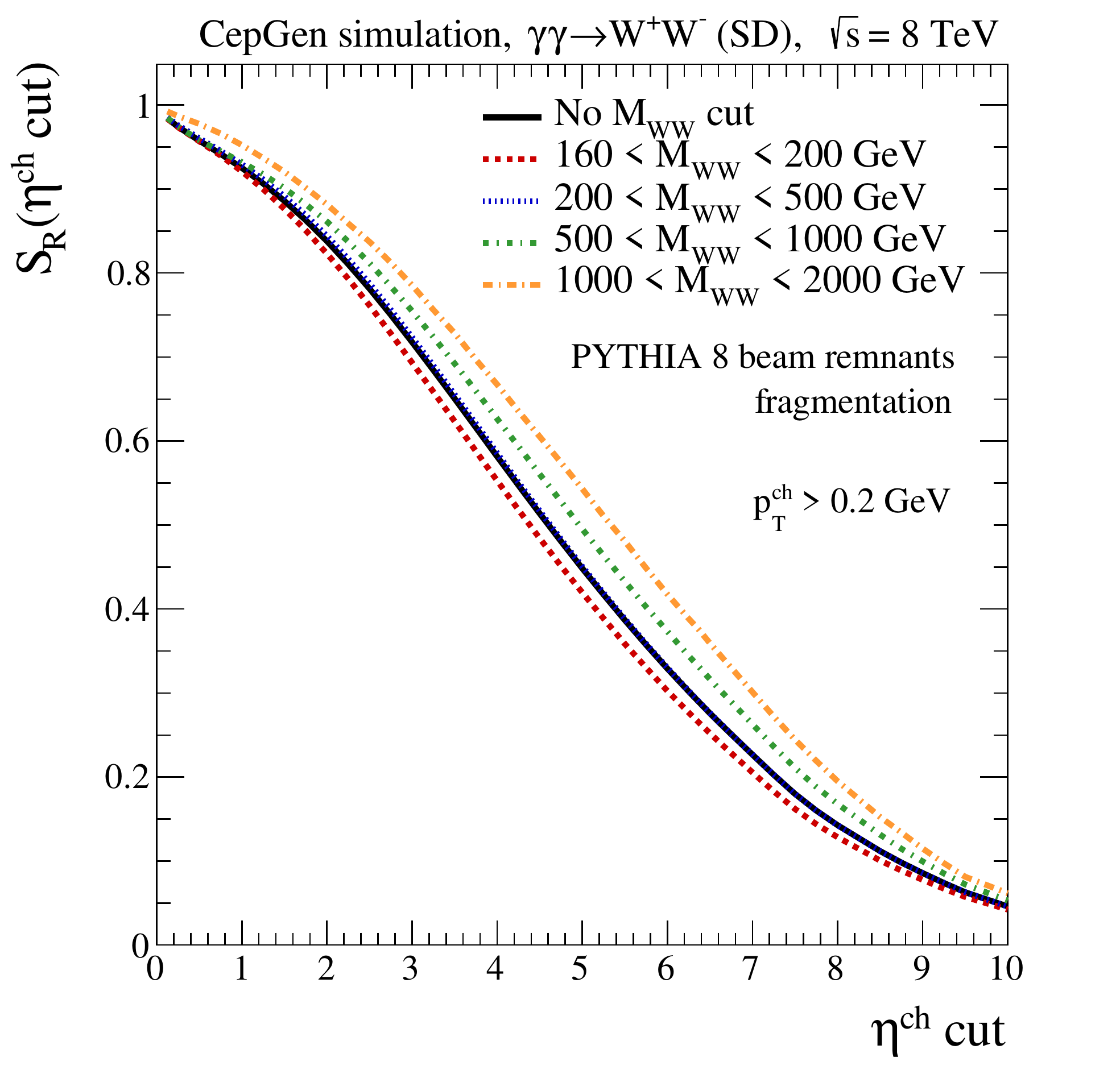}
  \includegraphics[width=.48\textwidth]{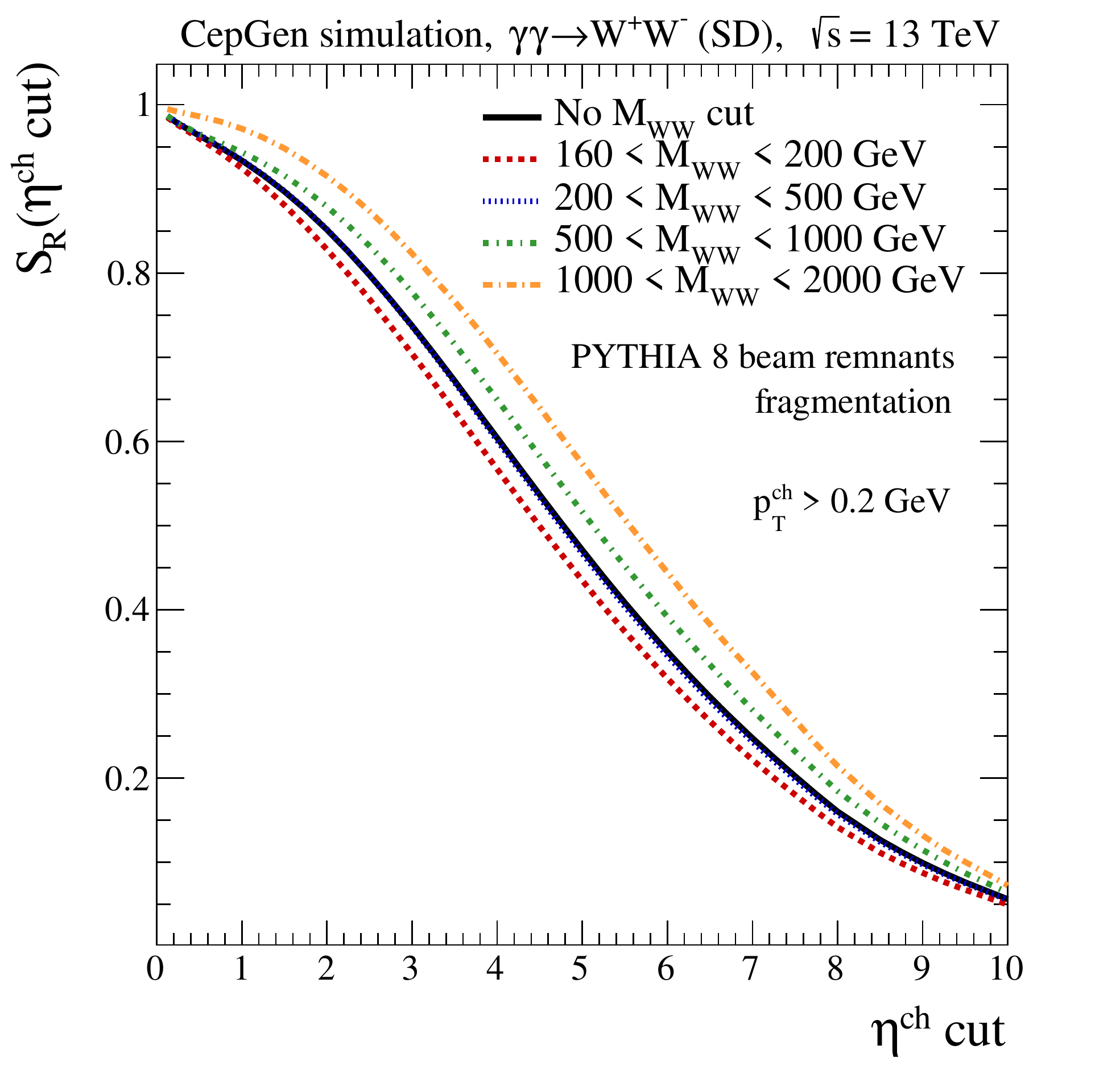}
  \caption{Gap survival factor for single dissociation as a function of
    the size of the pseudorapidity veto applied on charged particles emitted from proton
    remnants, for the
    diboson mass bins defined in the text and in the figures for $\sqrt{s} = 8~\TeV$ (left)
    and $13~\TeV$ (right).}
  \label{fig:surv_vs_etacut_sd}
\end{figure}

In Fig. \ref{fig:surv_vs_etacut_sd} we show the distribution in $\eta_{\rm cut}$
for single dissociative process. The numbers here are somewhat
larger than those shown in Fig. \ref{fig:surv_vs_etacut_dd}, consistently
with factorisation.
Detailed inspection shows \eqref{factorisation} holds for all $M_{WW}$ regions.

\begin{figure}
  \centering
  \includegraphics[width=.48\textwidth]{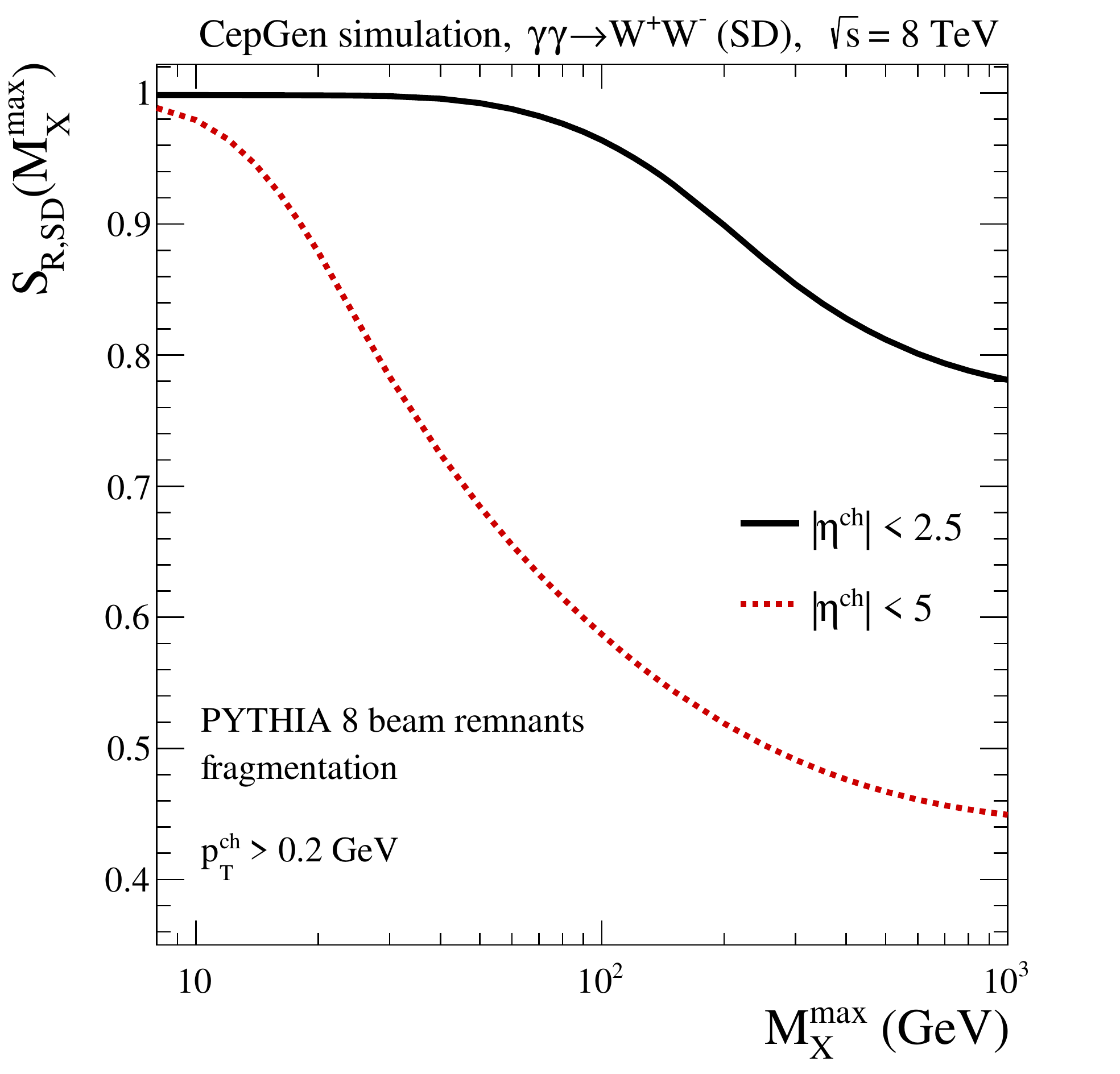}
  \includegraphics[width=.48\textwidth]{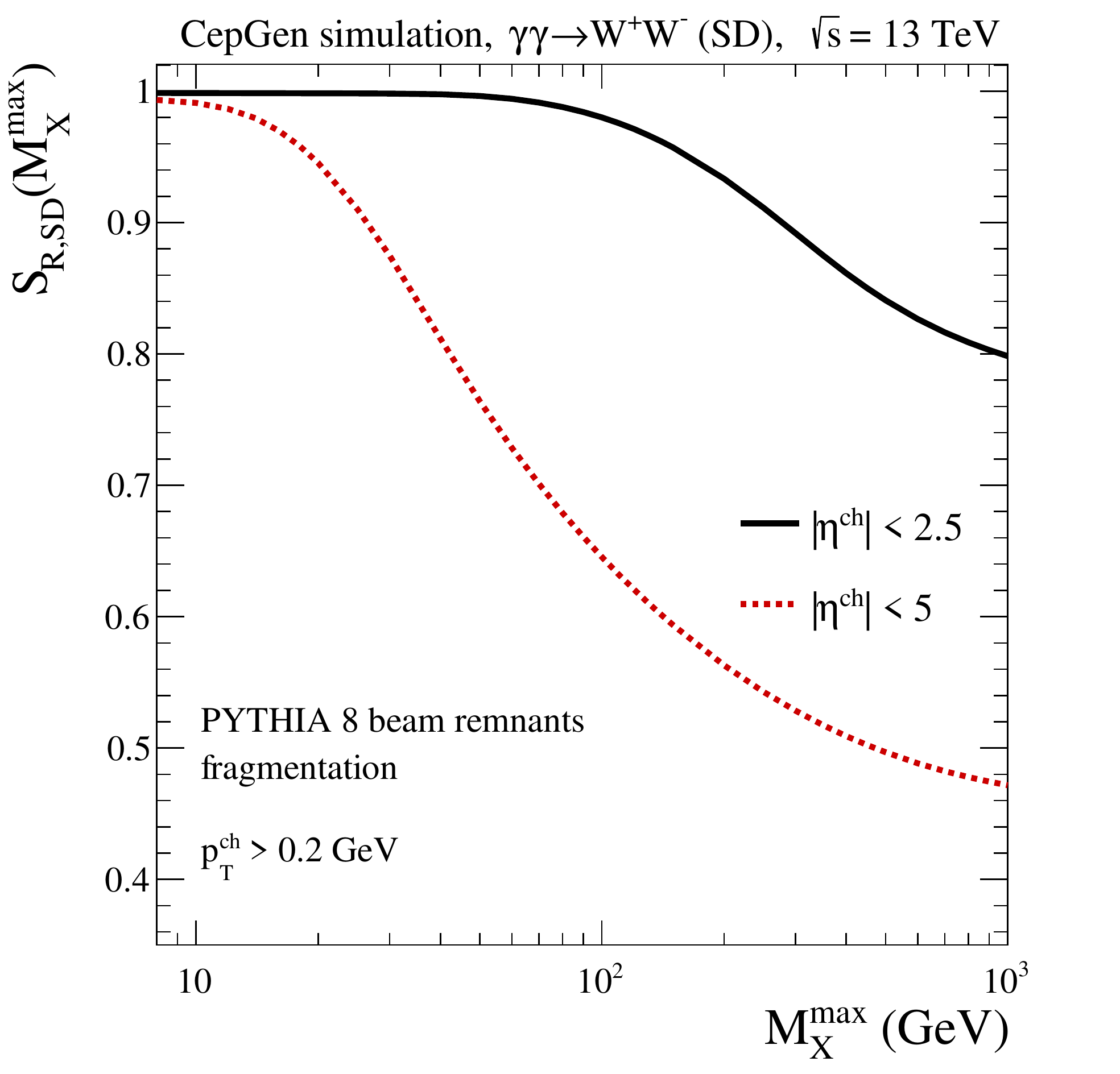}
  \caption{Rapidity gap survival factor for $|\eta^{\rm ch}|<2.5$ and $|\eta^{\rm ch}|<5$ as
a function of the upper limit set on $M_X$, the remnant system
invariant mass, for single dissociation.}
  \label{fig:surv_vs_mxmax}
\end{figure}

For later studies, the dependence of the rapidity gap survival factor
on the mass of the dissociated hadronic system may be interesting.
Corresponding results are shown in Fig. \ref{fig:surv_vs_mxmax}.
We observe that for an $\eta_{\rm cut}$ value of 2.5 the rapidity gap
survival factor $S_R$ stays very close to 1 for $M_X^{\rm max} < 100~\GeV$.
Increasing the mass of the dissociative system leads to graduate destroying
of the (pseudo)rapidity gap, arbitrarily fixed here to
$-2.5 < \eta < 2.5$ (ATLAS, CMS).

From Fig. \ref{fig:surv_vs_mxmax} one may infer which masses can be allowed
in the dissociation still ensuring the gap and avoiding a
more complicated Monte Carlo simulation of the remnant hadronisation.

The hadronisation part depends on the kinematics
of the centrally produced system, but otherwise is independent of the
quantum numbers of this system.
Hence, this method can be used to perform calculations for processes
for which there are no direct procedures to perform full Monte Carlo
simulations.

\begin{table}[tbp]
\centering
\begin{tabular}{|c|c|c|}
\hline
contribution               &  13 TeV  \\
\hline
        Inclusive                   &          \\

Exc.   & 0.11   \\
SD     & 0.44   \\
DD     & 0.47  \\
\hline
       $\eta_{\rm cut}$ = 6.5 ($\delta \sim 3$)      &               \\

Exc.   & 0.39   \\
SD     & 0.35   \\
DD     & 0.26  \\
\hline
     $\eta_{\rm cut}$ = 2.5 ($\delta \sim 7$)      &          \\

Exc.   & 0.15   \\
SD     & 0.58   \\
DD     & 0.19  \\
\hline

\end{tabular}
\caption{Relative contribution of exclusive (Exc.), single dissociative (SD) and
double dissociative (DD) contributions to photon-induced $W^+W^-$ production
at $\sqrt{s} = 13 \, \rm TeV$.
}
\label{table:relative_cont}
\end{table}

Let us come back to the ordering of different processes \eqref{inclusive_cs_ordering}.
In Table~\ref{table:relative_cont} we show
the relative contributions of exclusive (Exc.), single dissociative (SD)
and double dissociative (DD) processes for the inclusive case
(without gap requirement) as well as for $\eta_{\rm cut} = 2.5$  and
$\eta_{\rm cut} = 6.5$. Similar results are shown in Table 3 of
\cite{Harland-Lang:2016apc} (their $\delta =3 $ and $\delta =7$ correspond
to our $\eta_{\rm cut} = 6.5$ and $\eta_{\rm cut} =2.5$, respectively), including
effects of soft rescatterings in a simple two-channel eikonal model.
It can be seen that for $\eta_{\rm cut} = 2.5$ the results are in the same ballpark,
although after rescattering the exclusive fraction is larger than the DD one.
The main difference is for $\eta_{\rm cut} =6.5$, where in \cite{Harland-Lang:2016apc}
the DD contribution becomes entirely negligible. We note however that
generally large ranges of dissociative masses are relevant (see e.g. Fig~\ref{fig:surv_vs_mxmax}),
for which description a two-channel eikonal is not necessarily reliable.
The DD component in \cite{Harland-Lang:2016apc} is smaller than in
our case because of soft gap survival factor. 
The production associated with the two large masses $M_X$ and $M_Y$ for
DD is naturally associated with smaller impact parameter 
and consequently the $S_{soft}(DD)$ is rather small, smaller than 
e.g. for the SD components.

So far we have not included the soft gap survival factors.
They are relatively easy to calculate only for double elastic (DE)
contribution (see e.g \cite{Lebiedowicz:2015cea}).
For the ``soft'' gap survival factors we expect:
\begin{equation}
S_{soft}(DD) < S_{soft}(SD) < S_{soft}(DE) \; .
\end{equation}
Some estimates of phase space averaged values were presented in 
\cite{Harland-Lang:2016apc}.
A precise kinematics-dependent calculation of soft gap survival factor
requires further studies which go, however, beyond the scope
of the letter, devoted to remnant fragmentation.
We expect that the soft gap survival factors may violate the relation
$S_R(DD) = (S_R(SD))^2$ for the combined (remnant+soft) rapidity gap
survival factors.

\section{Conclusions}

In the present letter we have discussed the quantity called
``remnant gap survival factor'' for the $pp \to W^+ W^-$ reaction
initiated via photon-photon fusion.
We use a recent formalism developed
for the inclusive case \cite{Luszczak:2018ntp} which includes
transverse momenta of incoming photons.

First we have calculated the gap survival factor
for single dissociative process on the parton level.
In such an approach the outgoing parton (jet/mini-jet) is responsible
for destroying the rapidity gap. We have discussed the role of valence and
sea contributions.

Next the partonic formalism has been supplemented
here by including remnant fragmentation that can spoil the rapidity gap
usually used to select the subprocess of interest.
We quantify this effect by defining the remnant gap survival factor
which in general depends on the reaction, kinematic variables and
details of the experimental set-ups.
We have found that the hadronisation only mildly modifies
the gap survival factor calculated on the parton level.
This may justify approximate treatment of hadronisation of remnants.
We discus this dependence on invariant mass of the produced $W^+W^-$
central system. We find different values for double and single
dissociative processes. In general, $S_{R,DD} < S_{R,SD}$
and $S_{R,DD} \approx (S_{R,SD})^2$.
We expect that the factorisation observed here for
the remnant dissociation and hadronisation will be violated
when the soft processes are explicitly included.
Furthermore the larger $\eta_{\rm cut}$ (upper limit on charged
particles pseudorapidity), the smaller rapidity gap survival factor
$S_R$. This holds both for the double and the single dissociation.
Finally the effect becomes smaller for larger collision energies.
We have found that the crucial variable for $S_R$ is (are) masses
of the final hadronic remnant systems.

The present approach is a step towards a realistic modelling of
gap survival in photon induced interactions and definitely requires
further detailed studies and comparisons to the existing and future
experimental data.
In the present analyses we have neglected other effects such as soft
interactions or multiple-parton interactions
(see e.g. \cite{Khoze:2017sdd,Babiarz:2017jxc}).
More detailed studies including such effects in a consistent manner
will be given elsewhere.

\section*{Acknowledgements}

This study was partially supported by the Polish National Science Centre
grants DEC-2013/09/D/ST2/03724 and
DEC-2014/15/B/ST2/02528 and by the Center for Innovation and
Transfer of Natural Sciences and Engineering Knowledge in Rzesz{\'o}w.
We thank the financial support from the grant of C. Royon as
a Foundation Distinguished Professor. M.{\L}. thanks CERN for the hospitality,
where this work was finalised.
We are indebted to R. Staszewski for a helpful discussion.
L.F. thanks T. Sj\"{o}strand for useful discussions in
the \pythia implementation of the beam remnants fragmentation.
We are indebted to Valery Khoze and Lucian Harland Lang for a discussion
on soft rapidity gap survival factors.

\bibliographystyle{apsrev}
\bibliography{pp_WW_gapsurvival}


\end{document}